\documentclass[11pt,a4paper,twoside]{article}

\usepackage{amssymb}
\usepackage{amsmath}
\usepackage{bm}
\usepackage{graphicx}
\usepackage{subfigure}
\usepackage{hyperref}
\usepackage{color}
\usepackage[font=small,labelfont=bf]{caption}

\usepackage{booktabs}
\usepackage{array}
\usepackage{enumitem}
\usepackage{cite}
\usepackage[affil-it,auth-lg]{authblk}
\usepackage{hyperref}
\usepackage{eso-pic}

\let\OLDthebibliography\thebibliography
\renewcommand\thebibliography[1]{
	\OLDthebibliography{#1}
	\setlength{\parskip}{0pt}
	\setlength{\itemsep}{0pt plus 0.3ex}
}

\begin{document}
	
	\title{Gravitational instability caused by the weight of heat}
	\author[1]{Zacharias Roupas}
	\affil[1]{Centre for Theoretical Physics, The British University in Egypt, Sherouk City 11837, Cairo, Egypt} 
	\date{\vspace{-5ex}}
	\maketitle

\begin{abstract}
Thermal energy points toward a disordered, completely uniform state acting counter to gravity's tendency to generate order and structure through gravitational collapse. It is, therefore, expected to contribute to the stabilization of a self-gravitating, classical ideal gas over collapse. However, I identified an instability that always occurs at sufficiently high energies: the high-energy or relativistic gravothermal instability. I argue here that this instability presents an analogous core--halo structure as its Newtonian counterpart, the Antonov instability. The main difference is that in the former case the core is dominated by the gravitation of thermal energy and not rest mass energy. A relativistic generalization of Antonov's instability---the low-energy gravothermal instability---also occurs. The two turning points, which make themselves evident as a double spiral of the caloric curve, approach each other as relativistic effects become more intense and eventually merge in a single point. Thus, the high and low-energy cases may be realized as two aspects of a single phenomenon---the gravothermal instability---which involves a core--halo separation and an intrinsic heat flow. Finally, I argue that the core formed during a core-collapse supernova is subject to the relativistic gravothermal instability if it becomes sufficiently hot and compactified at the time of the bounce. In this case, it will continue to collapse towards the formation of a black hole.
\end{abstract}

\section{Introduction}

In a seminal paper, Tolman~\cite{Tolman:1930} discovered that the local temperature of a self-gravitating system is not constant in equilibrium if general relativity is taken into account. Quoting his words, ``heat has weight''. Thermal energy rearranges itself in order to balance its own gravitational attraction~\cite{Tolman-Ehrenfest:1930}. This results to a local temperature gradient at~equilibrium. 

Many years later, in~another seminal paper, Antonov~\cite{antonov} discovered that in the Newtonian limit, and for an ideal gas, there exists a minimum energy below which no stable equilibria exist under conditions of constant energy. Lynden-Bell and Wood~\cite{lbw} described the mechanism underlying this Antonov instability, which they named gravothermal catastrophe. The~system becomes unstable as we move along the series of equilibria from stable states with negative specific heat, to the unstable branch with positive specific heat. This instability occurs as the energy of the system is lowered, which causes the contraction of the central parts in an attempt to generate a sufficient pressure~gradient.

{In} \cite{Roupas_CQG_RGI_2015} I raised the question of what happens if we follow the opposite direction in the caloric curve, i.e.,~we move along higher and higher energies until relativistic effects start to become relevant. The aim of this paper is to investigate the competition of thermal energy and gravity over the stability of the system, and therefore I maintain the assumption of ideal gas neglecting the complexities which a more involved equation of state would introduce. Under~this perspective, the~statistical aspect of the ideal gas equation of state is emphasized. It is viewed as the statistical distribution of microstates unconstrained by interactions, with the particles being entirely independently distributed among the various states. How is this distribution affected by the presence of the negative gravitational potential generated by phase-space as in General Relativity and not solely configuration space as in the Newtonian limit? Will the weight of heat of Tolman somehow manifest itself?

\subsection*{A Thought~Experiment}
 Imagine a spherical box containing a self-gravitating ideal gas that cannot exchange energy with the environment outside the box. Assume its total gravitational plus thermal energy to be negative, so that it is a system bound by self-gravity and not by the external walls; the~latter are incorporated only to prevent the evaporation of the system at all timescales. Suppose the system achieves an equilibrium state and that we start slowly to increase the radius of the sphere. The expansion cools the system down. The density profile becomes steeper as more mass tends towards the center due to the reduced ability of heat to counterbalance~gravity.

If we continue to expand the sphere we reach a threshold (point $B$ in Figure~\ref{fig:spirals}c), beyond which the temperature rises during expansion. The system has attained negative heat capacity. It has become so condensed that the central parts are bound primarily by self-gravitation and not by the outer parts or the box. In~this case, condensation (note that while the box expands the central parts condense) causes heating because of negative gravitational energy. This is dictated by virial theorem, or~may be understood very simply in the case of a single body moving in a central potential. Equilibria closer to the center correspond to larger orbital velocities. This negative specific heat branch is stable if the energy of the box is~conserved.

However, as~we continue to move along this negative specific heat branch expanding the sphere, we reach a second threshold (point $A$ in Figure~\ref{fig:spirals}c) beyond which the system attains again positive specific heat, but~now becomes unstable. Lynden-Bell and Wood~\cite{lbw} explained this as follows. The~self-gravitating core decouples from the outer parts attaining negative-specific heat. As the total specific heat of the core and the outer parts---the halo---is positive, a temperature gradient
from the core to the halo cannot be reversed\footnote{Here, and in the following, we assume spherical symmetry and therefore by ``gradient'' is meant a radial derivative.}. A~runaway effect of heat transfer takes place. Both the core and the halo become hotter and a temperature equalization is impossible. This is the Antonov instability or gravothermal catastrophe. 
{{Note that} if we continue to expand the sphere, we will reach another threshold beyond which stable equilibria do exist at sufficiently big radii such that the dark energy becomes relevant. Entropy maxima will be restored due to the stabilizing, repulsive nature of dark energy~\cite{Axenides_2012PhRvD..86j4005A,Axenides_2013NuPhB.871...21A,Roupas_2014PhRvD..89h3002R}.} 

Now, imagine that we compress the sphere back beyond the first threshold (starting from $A$ and moving towards $B$ in Figure~\ref{fig:spirals}c) and enter again the stable, positive specific heat branch of series of equilibria. As~the sphere is compressed, the~gas heats up and becomes more uniform. The~mass density profile flattens, tending to a constant density, uniform state, as~expected. However, surprisingly, we reach a point (point ${\Sigma}$ in Figure~\ref{fig:spirals}c,d) beyond which any compression causes a steepening of the mass density profile. Relativity has started to become important. The~mass density now includes the thermal mass of random movement of the particles. During~compression, this thermal mass concentrates to the center to generate a density gradient to counterbalance its own gravitational attraction, similarly to rest mass did during expansion. At~some point, we reach another threshold (point ${\Gamma}$ in Figure~\ref{fig:spirals}c,d), beyond which the system starts to cool down during compression. The system has attained again negative heat capacity. However, by~``cooling down'' we do not refer to the local temperature, but to the quantity that is conjugate of energy in general relativity and is uniform in equilibrium. Locally, the~temperature of the core continues to rise. The~thermal core becomes bound by its own gravity (likewise the rest mass core did in the previous case of expansion). Finally, we reach the point (called ${\Delta}$ in Figure~\ref{fig:spirals}c,d) when specific heat becomes positive again and another instability sets in, the~high energy gravothermal instability. The~core decouples from the outer regions---the halo---and collapses due to a runaway heat transfer from the core to the halo, like in the low-energy case. However, now, responsible for the decoupling, and the self-gravitation of the core is thermal mass and not rest mass. 
We will further quantify and analyze this thought experiment in Section~\ref{sec:RGI}.

\section{Hydrostatic Equilibrium of Relativistic Self-Gravitating Ideal~Gas}

\subsection{Tolman--Oppenheimer--Volkoff~Equation}

The equation that describes hydrostatic equilibrium in General Relativity is called the Tolman--Oppenheimer--Volkoff equation (TOV equation). It may be derived from Einstein's equations for a perfect gas~\cite{1934rtc..book.....T,1939PhRv...55..374O} and expresses a maximum entropy state~\cite{swz,Gao:2011hh,Roupas_2013CQG,Roupas_2015CQG_32k9501R,Fang_2014PhRvD..90d4013F,2015CQGra..32r5011S}. 

Let us denote $P(r)$, $\rho(r)$, and $T(r)$ as the pressure, the~total mass-energy density at $r$, and~the temperature measured by a static observer at $r$, respectively.
For an equation of state of the form
\begin{equation}
P(r) = P(\rho(r),T(r)),
\end{equation}
the hydrostatic equilibrium is expressed by three equations:
\begin{align}
		\label{eq:TOV}
	\frac{dP}{dr} &= -(\rho + \frac{P}{c^2})\left( \frac{G\hat{M}(r)}{r^2} + 4\pi G \frac{P}{c^2}r\right)
	\left( 1 - \frac{2G\hat{M}(r)}{rc^2} \right)^{-1}, \\
	\label{eq:Tprime}
	\frac{dT}{dr} &= \frac{T}{P + \rho c^2} \frac{dP}{dr}, \\
	\label{eq:massd}
	\frac{d\hat{M}}{dr} &= 4\pi r^2 \rho(r),
\end{align}
where $\hat{M}(r)$ denotes the total mass-energy, including that of gravitation, enclosed within $r$. $M$ denotes the total mass-energy of the system with radius $R$
\begin{equation}\label{eq:total_mass}
M \equiv \hat{M}(R) = \int_0^R \rho(r) 4\pi r^2 dr.
\end{equation}

 Equation~(\ref{eq:TOV}) is TOV equation, and Equation~(\ref{eq:Tprime}) is many times referred to as a Tolman relation. It may also be   expressed in the form 
$
T(r)\sqrt{\xi^\mu\xi_\mu} = const.
$, 
where $\xi^\mu$ is the time-like killing vector. Given the equation of state, the~system of Equations~(\ref{eq:TOV})--(\ref{eq:massd}) may be integrated to obtain the pressure, density, and temperature profiles at~equilibrium.

Equation~(\ref{eq:Tprime}) encapsulates the so-called Tolman--Ehrenfest effect~\cite{Tolman--Ehrenfest:1930}. This effect accounts for the fact that, inside a gravitational field, not only rest mass, but also ``heat'', in the sense of random kinetic energy, rearranges itself to counterbalance its own gravitational attraction. As~Tolman puts it, ``heat has weight''. On~the nature of the Tolman--Ehrenfest effect and the weight of heat, one may also consult~\cite{Balazs_1965Phy....31..222B,Ebert_1973GReGr...4..375E,Stachel_1984FoPh...14.1163S,Rovelli_2011CQGra..28g5007R,Santiago_2018arXiv180304106S} and the references~therein.

We may gain further insight into this effect if we consider the limits $P\ll \rho c^2$ and $G\hat{M} \ll r c^2$, in Equation (\ref{eq:Tprime}), which give
\begin{equation}\label{eq:NewTol}
	\frac{1}{T}\frac{dT}{dr} = \frac{g}{c^2},
\end{equation}
where $g = -G\hat{M}/r^2$ denotes the Newtonian gravitational field. Let us derive it from the maximum entropy principle. Assume that a quantity of heat $|dE_1|$ flows from the subsystem $1$ to a subsystem $2$ at lower gravitational potential by $\Delta \phi$. The~energy $dE_2$ received by the second subsystem is not equal to $-dE_1$, but~equal to $dE_2 = - (dE_1 + m_h \Delta \phi)$, where $m_h = |dE_1|/c^2$ is the gravitational mass corresponding to the transferred heat. Now, assuming that the two systems achieve equilibrium, the~entropy is $dS = dS_1+dS_2 = 0$, which, after differentiating by $dE_2$ and using $1/T = dS/dE$, gives
\begin{equation}
	\frac{dS_1}{dE_1} = \frac{dS_2}{dE_2} \left( 1 - \frac{\Delta \phi}{c^2}\right) 
	\Rightarrow
	\frac{\Delta T}{T} = -\frac{\Delta \phi}{c^2},
\end{equation}
which expresses Equation~(\ref{eq:NewTol}). Evidently, the~temperature gradient is a result of the ``mass of heat'' $m_h = |dE_1|/c^2$.

\subsection{Equation of~State}

Let us briefly review the equations that describe the relativistic ideal gas. The~interested reader may further consult the work in~\cite{chandrabook}. 
For a quantum ideal gas, the~one-particle energy distribution is given by the
Fermi--Dirac or 
Bose--Einstein distributions for fermions or bosons, respectively:
\begin{equation}\label{eq:quantum_dis}
	g(\epsilon) = \frac{1}{e^{\beta(\epsilon - \mu)}\pm 1} \;,\;
	\left\lbrace
	\begin{array}{l}
		(+)\;\mbox{for fermions} \\[2ex]
		(-)\;\mbox{for bosons}
	\end{array}
	\right.
\end{equation}
where $\epsilon$ is the energy per particle, including rest mass in the relativistic case; $\mu$ the chemical potential; and $\beta = 1/kT$ is the inverse temperature. We focus on fermions. Substituting the relativistic definition of energy, 
$
	\epsilon = \sqrt{m^2c^4 + p^2 c^2},
$
where $m$ is the mass of one particle and $p$ its momentum, and~applying the 
Juettner transformation, 
$
	\frac{p}{mc} = \sinh\theta,
$
distribution (\ref{eq:quantum_dis}) may be written in terms of $\theta$:
\begin{equation}\label{eq:fb_dis}
	g(\theta) = \frac{1}{e^{b \cosh\theta - \alpha} + 1},\quad \text{where }\;
	b = \frac{mc^2}{kT },\;
	\alpha = \frac{\mu}{kT}. 
\end{equation}

Using distribution (\ref{eq:fb_dis}), one may show~\cite{chandrabook} that the pressure, $P$; number density, $n$; and total
mass-energy density, $\rho$, may be written as
\begin{align}
\label{eq:P_Q} 	
&P = \frac{4\pi g_s m^4 c^5}{3h^3}\int_0^\infty \frac{\sinh^4\theta d\theta}{e^{b \cosh\theta - \alpha}+1}
\\
\label{eq:rho_Q}
&\rho = \frac{4\pi g_s m^4 c^3}{h^3}\int_0^\infty \frac{\sinh^2\theta \cosh^2\theta  d\theta}{e^{b \cosh\theta - \alpha}+1}
\\
\label{eq:n_Q}
&n = \frac{4\pi g_s m^3 c^3}{h^3}\int_0^\infty \frac{\sinh^2\theta \cosh\theta  d\theta}{e^{b \cosh\theta - \alpha}+1}
,
\end{align}
where $h$ is Planck constant and $g_s$ is the degeneracy of the quantum state, for~example, $g_s = 2$ for electrons and neutrons that have spin $1/2$. In~the classical limit, $\beta\epsilon-\alpha\gg 1$, we get
\begin{equation}
	g(\epsilon) \rightarrow e^{-\beta \epsilon + \alpha}.
\end{equation}

Equations~(\ref{eq:P_Q})--(\ref{eq:n_Q}) become
\begin{align}
\label{eq:P_clasB} 	
&P = \frac{4\pi g_s m^4 c^5}{h^3}e^{\alpha}\frac{K_2(b)}{b^2}
\\
\label{eq:rho_clasB}
&\rho = \frac{4\pi g_s m^4 c^3}{h^3}e^{\alpha}\frac{K_2(b)}{b}(1+\mathcal{F}(b))
\\
\label{eq:n_clasB}
&n = \frac{4\pi g_s m^3 c^3}{h^3}e^{\alpha}\frac{K_2(b)}{b}.
\end{align}
where
\begin{equation}\label{eq:F}
		 \mathcal{F}(b) = \frac{K_1(b)}{K_2(b)} + \frac{3}{b} - 1
\end{equation}
and $K_\nu(b)$ are the modified Bessel functions
\begin{equation}\label{eq:bessel}
	K_\nu(b) = \int_0^{\infty} e^{-b \cosh\theta}\cosh(\nu\theta)d\theta.	
\end{equation}

We used the recursive relations
\begin{equation}\label{eq:recur}
	K_{\nu+1}(b) - K_{\nu-1}(b) = \frac{2\nu}{b}K_\nu(b).
\end{equation}

Equations~(\ref{eq:P_clasB})--(\ref{eq:n_clasB}) give the equation of state of the relativistic classical ideal gas:
\begin{equation}\label{eq:clas_eos}
	P = \frac{n m  c^2}{b} \; \mbox{, or~equivalently}\;
	P = \frac{\rho c^2}{b(1+\mathcal{F})}.
\end{equation}

\section{Gravothermal~Instability}\label{sec:RGI}

\begin{figure}[tb]
	\centering
	\subfigure[Low-energy gravothermal instability]{\label{fig:spiral_high}
		\includegraphics[scale = 0.4]{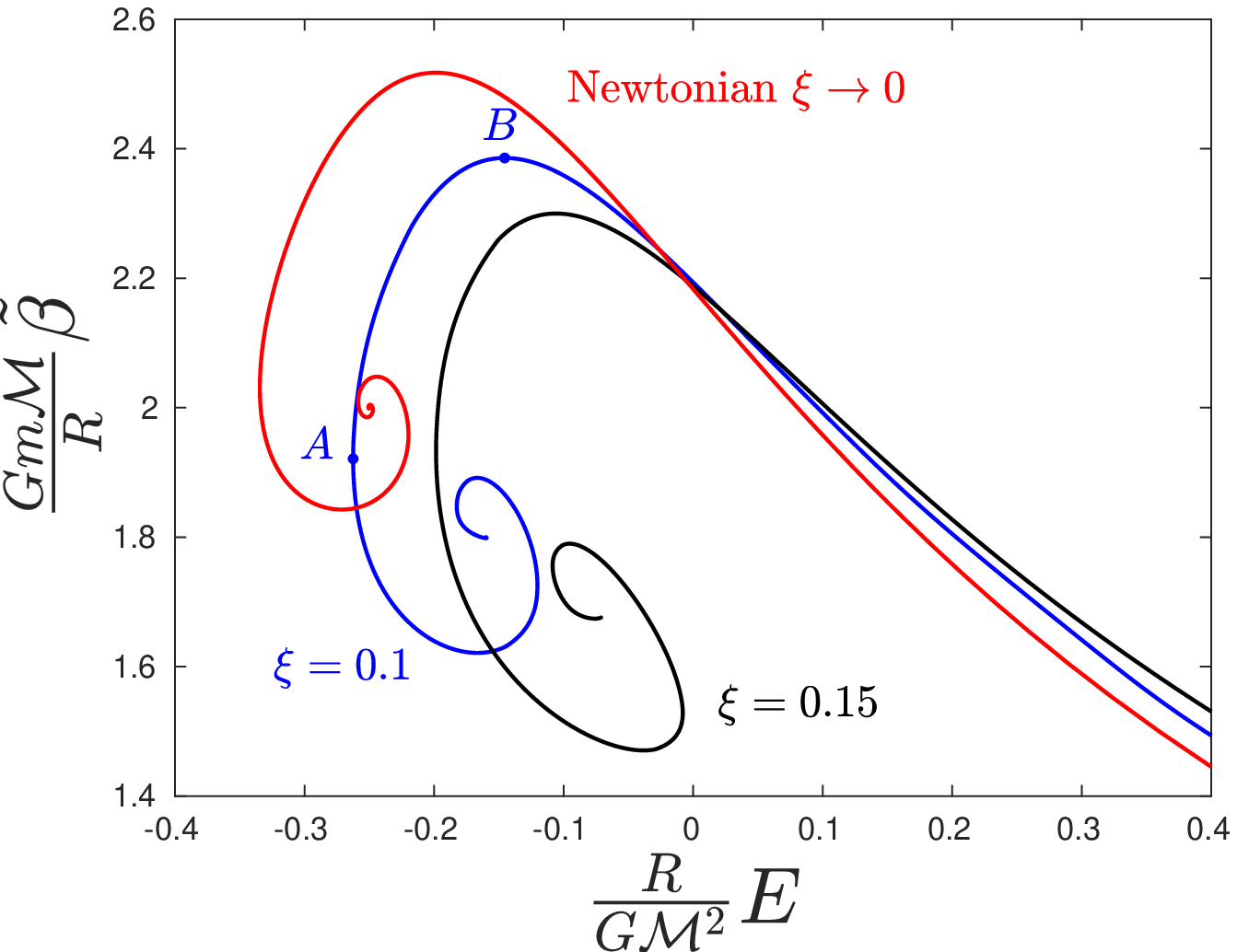}  }
	\subfigure[High-energy gravothermal instability]{\label{fig:spiral_low}
		\includegraphics[scale = 0.4]{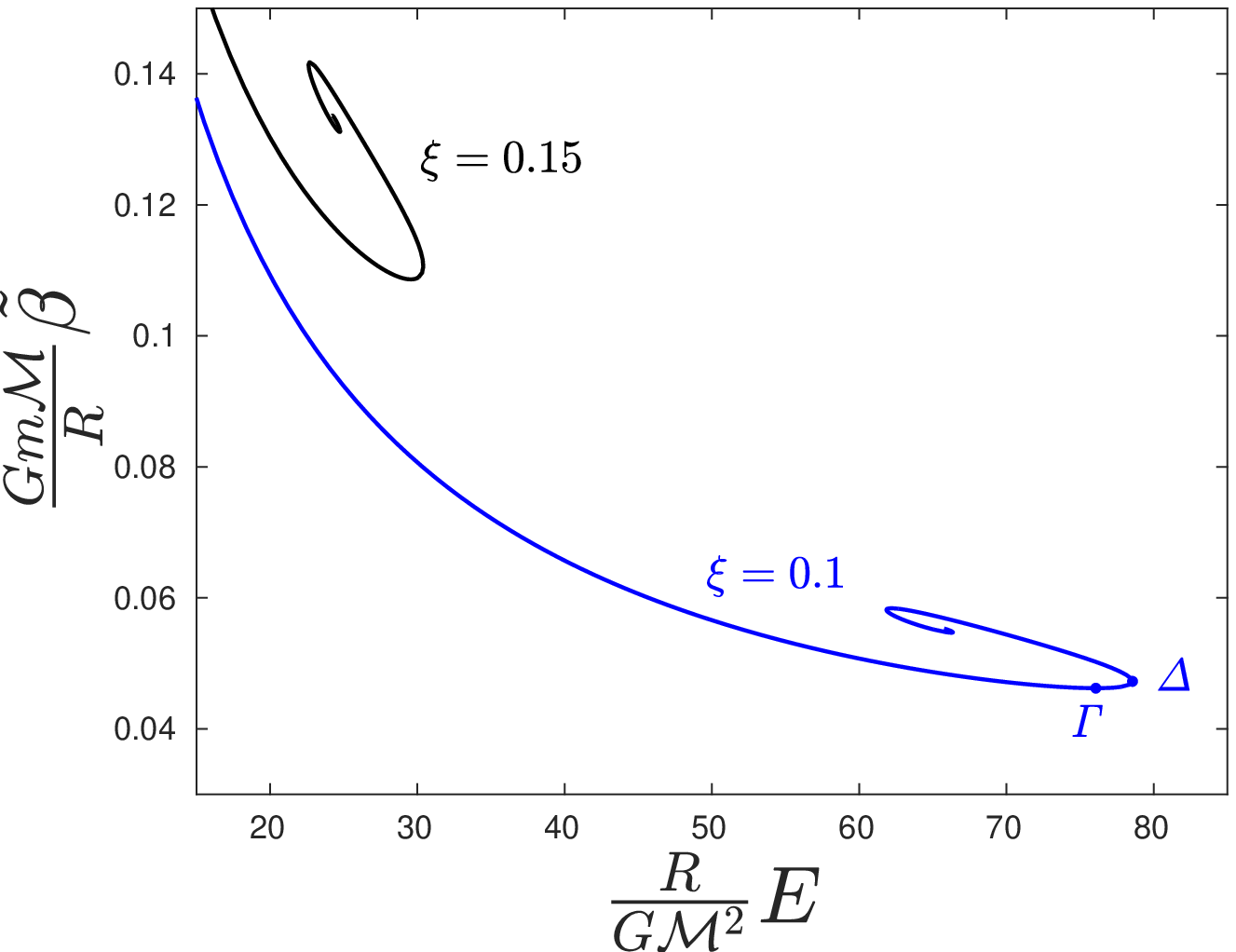}  }
	\\
	\subfigure[Low energy, $\xi=0.1$]{ \label{fig:CV_high-spiral}
		\includegraphics[scale = 0.4]{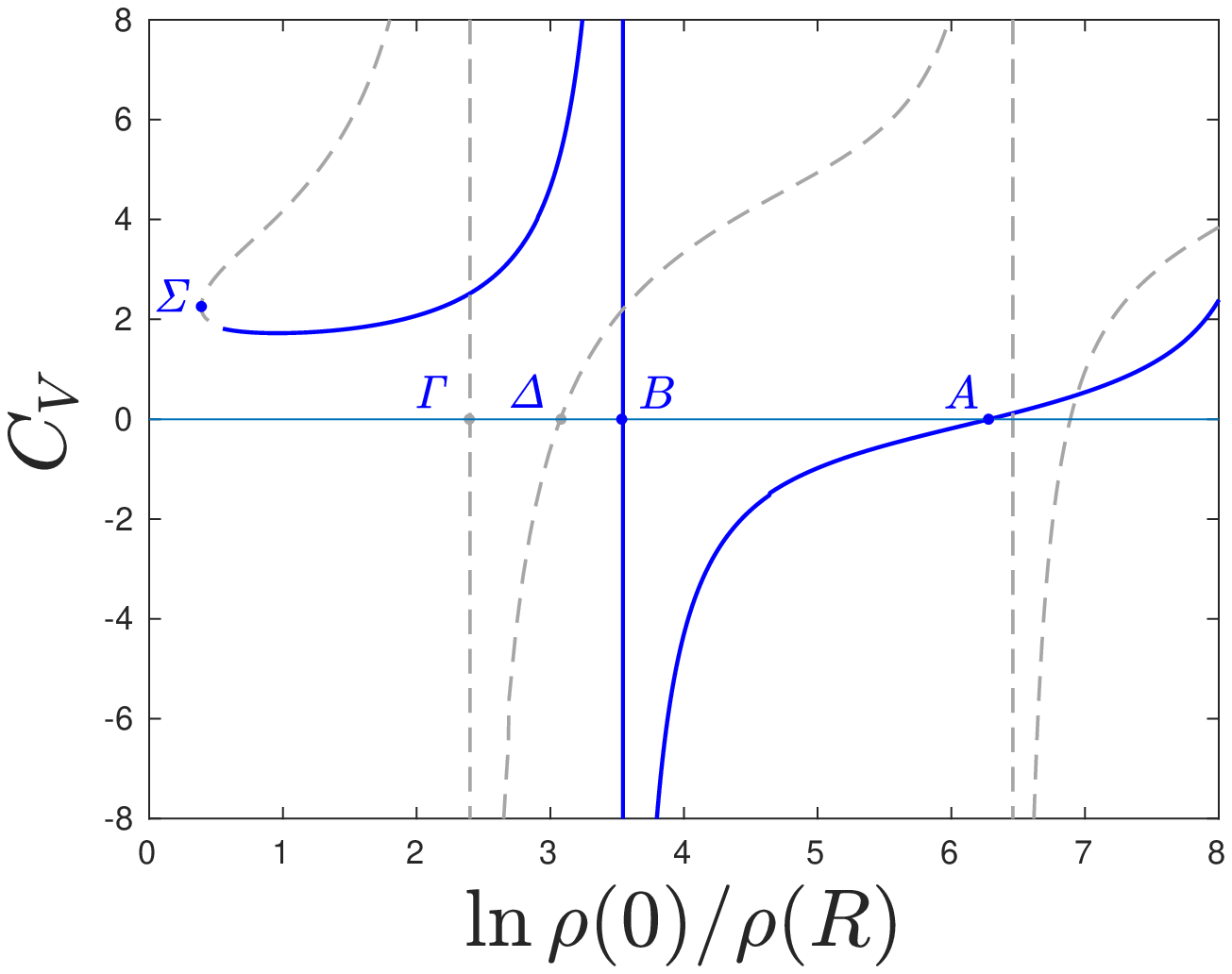}  }
	\subfigure[High energy, $\xi=0.1$]{ \label{fig:CV_low-spiral}
		\includegraphics[scale = 0.4]{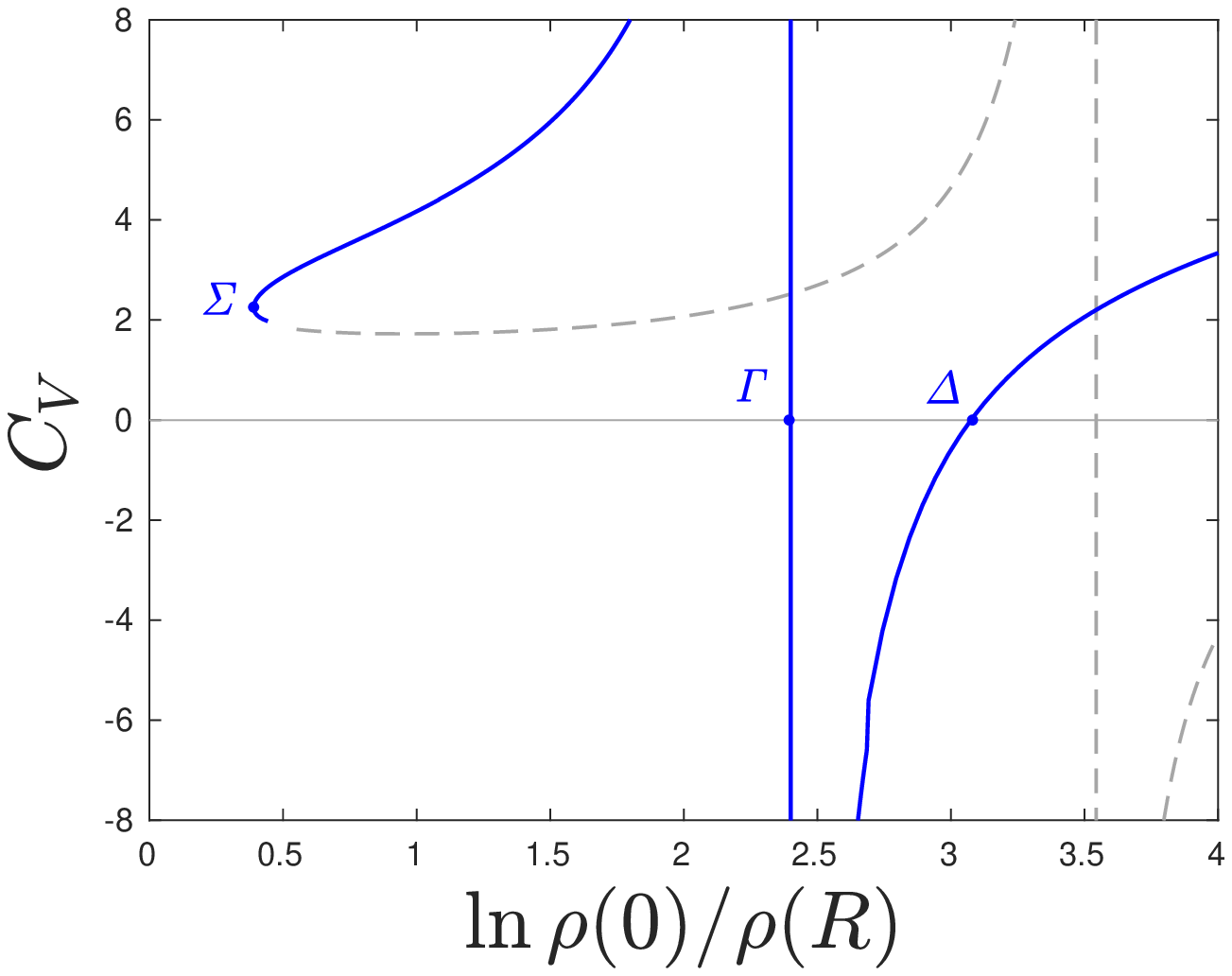}  }
	\caption{Upper panels: The caloric curves $\tilde{\beta} = \tilde{\beta}(E)$ for fixed rest compactness $\xi = 2G\mathcal{M}/Rc^2$, $\mathcal{M} = mN$. Both $N$ and $R$ may be regarded as constant in these plots. We denote $E = Mc^2 - \mathcal{M}c^2$ the gravothermal energy of the system. (\textbf{a}) The low-energy gravothermal instability for the values of rest compactness, $\xi=0.1$ and $0.15$, together with the Newtonian limit $\xi\rightarrow 0$, corresponding to Antonov instability. The~instability occurs at minimum energy (point $A$ for $\xi=0.1$). (\textbf{b}) The high-energy or relativistic gravothermal instability, for~the values of rest compactness $\xi=0.1$, $0.15$. The~instability occurs at maximum energy (point ${\Delta}$ for $\xi=0.1$). In~the Newtonian limit, $\xi \rightarrow 0$ it is  $E_{\Delta}\rightarrow \infty$. The~two spirals, at~low and high energies are connected with a stable series of equilibria, not shown here, but~depicted in {Figure}~\ref{fig:Spiral_double} for $\xi=0.25$. 
		Lower panels: The specific heat w.r.t. the density contrast for $\xi=0.1$. Both panels (\textbf{c},\textbf{d}) depict the same diagram. In~panel (c), the low-energy gravothermal instability is highlighted. The point ${\Sigma}$ denotes the threshold, beyond~which thermal energy takes over and any increase of energy causes an increase and not decrease of density contrast (the system becoming less homogeneous). Both low- and high-energy instabilities, at~points $A$, ${\Delta}$, respectively, occur as we move from the stable negative heat branch to positive specific heat, designating a core--halo structure.   
		\label{fig:spirals}}
\end{figure}

Let us use the following dimensionless variables to solve the TOV equation
\begin{equation}
	x = \frac{r}{r_\star},\;
	u = \frac{\hat{M}}{M_\star},\;
	r_\star = \left( \frac{4\pi G}{c^2}\rho_0\right)^{-\frac{1}{2}},\;
	M_\star = r_\star\frac{c^2}{G},\;
	\psi = \ln\frac{b}{b_0},\;
	\bar{\rho} = \frac{\rho}{\rho_0}.
\end{equation}

We denote $m$ as the rest mass of one particle, $\rho_0$ as the total mass-energy density at the origin, and $b_0 = b(r=0)$. TOV equation~(\ref{eq:TOV}) and mass equation~(\ref{eq:massd}) become, by use of Equations (\ref{eq:Tprime}), (\ref{eq:P_clasB}), and (\ref{eq:rho_clasB}),
\begin{align}
	\label{eq:TOV_nonD}
&\frac{d\psi(x)}{dx} = \left( \frac{u(x)}{x^2} + \frac{\bar{\rho}(x)}{b(x)(1+\mathcal{F}(b(x)))}x\right)
	\left( 1 - \frac{2 u(x)}{x} \right)^{-1}, \\
\label{eq:massd_nonD}
&\frac{d u(x)}{d x} = \bar{\rho}(x) x^2,
\end{align}
where $\mathcal{F}(b)$ is given by Equation~(\ref{eq:F}) and
\begin{equation}
	\bar{\rho} = \frac{K_2(b)(1+\mathcal{F}(b))}{b}/\left( \frac{K_2(b_0)(1+\mathcal{F}(b_0)}{b_0} \right)
\end{equation} 

This forms the system to be solved with initial conditions $\psi(0) = 0$, $u(0)=0$. To generate the caloric curves, the~boundary radius of integration
\begin{equation}
	z = \frac{R}{r_\star}
\end{equation}
is chosen for each $b(0)=b_0$, such that the compactness of rest mass, hereafter known as ``rest compactness'', to distinguish from the usual compactness $2GM/Rc^2$,
\begin{equation}
	\xi \equiv \frac{2G\mathcal{M}}{Rc^2} = \frac{2}{z} \int_0^z \frac{(n(r)/n_0)}{1+\mathcal{F}(b_0)} \left( 1 - \frac{2 u}{x^2} \right)^{-\frac{1}{2}} x^2 dx
\end{equation}
is kept constant. The~rest compactness controls the intensity of relativistic effects. 
We denote the total rest mass
\begin{equation}
	\mathcal{M} = mN
\end{equation}
and the rest mass-energy density is given by use of (\ref{eq:n_clasB}):
\begin{equation}\label{eq:n_nonD}
	\frac{\rho_\text{rest}(r)}{\rho_{\text{rest},0}} \equiv \frac{mn(r)}{mn_0} = \frac{K_2(b(r))}{b(r)}/\left( \frac{K_2(b_0)}{b_0}\right).
\end{equation}

The pressure and thermal mass-energy density $\rho_\text{therm} = \rho - \rho_\text{rest}$ may be calculated from the~expressions
\begin{align}
\label{eq:P_nonD} 	
&\frac{P}{P_0} = \frac{K_2(b)}{b^2}/\left( \frac{K_2(b_0)}{b_0^2}\right),
\\
\label{eq:rho_th_nonD}
&\frac{\rho_\text{therm}}{\rho_{\text{therm},0}} = \frac{K_2(b)\mathcal{F}(b)}{b}/\left( \frac{K_2(b_0)\mathcal{F}(b_0)}{b_0}\right).
\end{align}

\begin{figure}[tb]
	\begin{center}
		\subfigure[Low-energy gravothermal instability]{ \label{fig:rho_upper-spiral}
			\includegraphics[scale = 0.4]{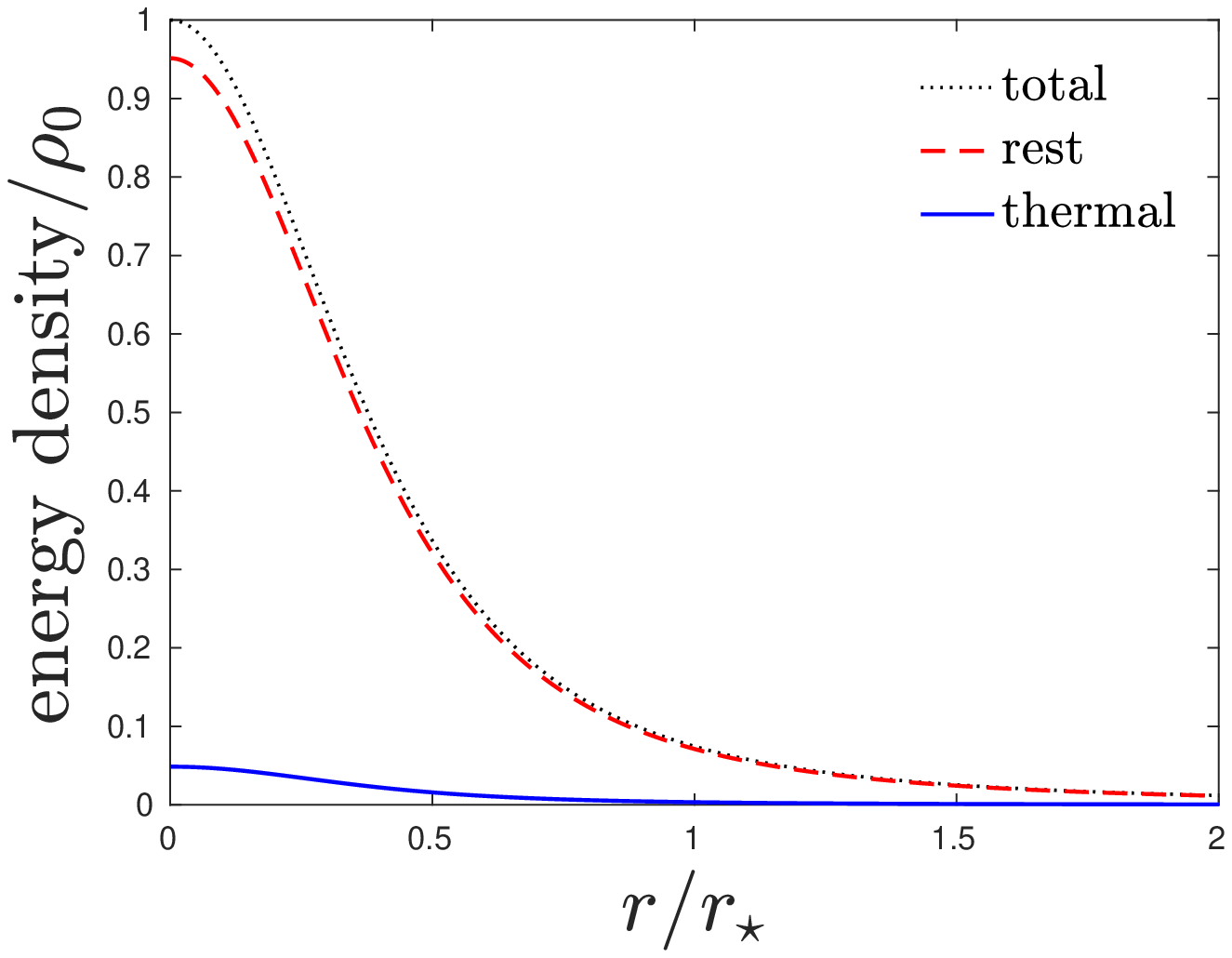}  }
		\subfigure[High-energy gravothermal instability]{ \label{fig:rho_lower-spiral}
			\includegraphics[scale = 0.4]{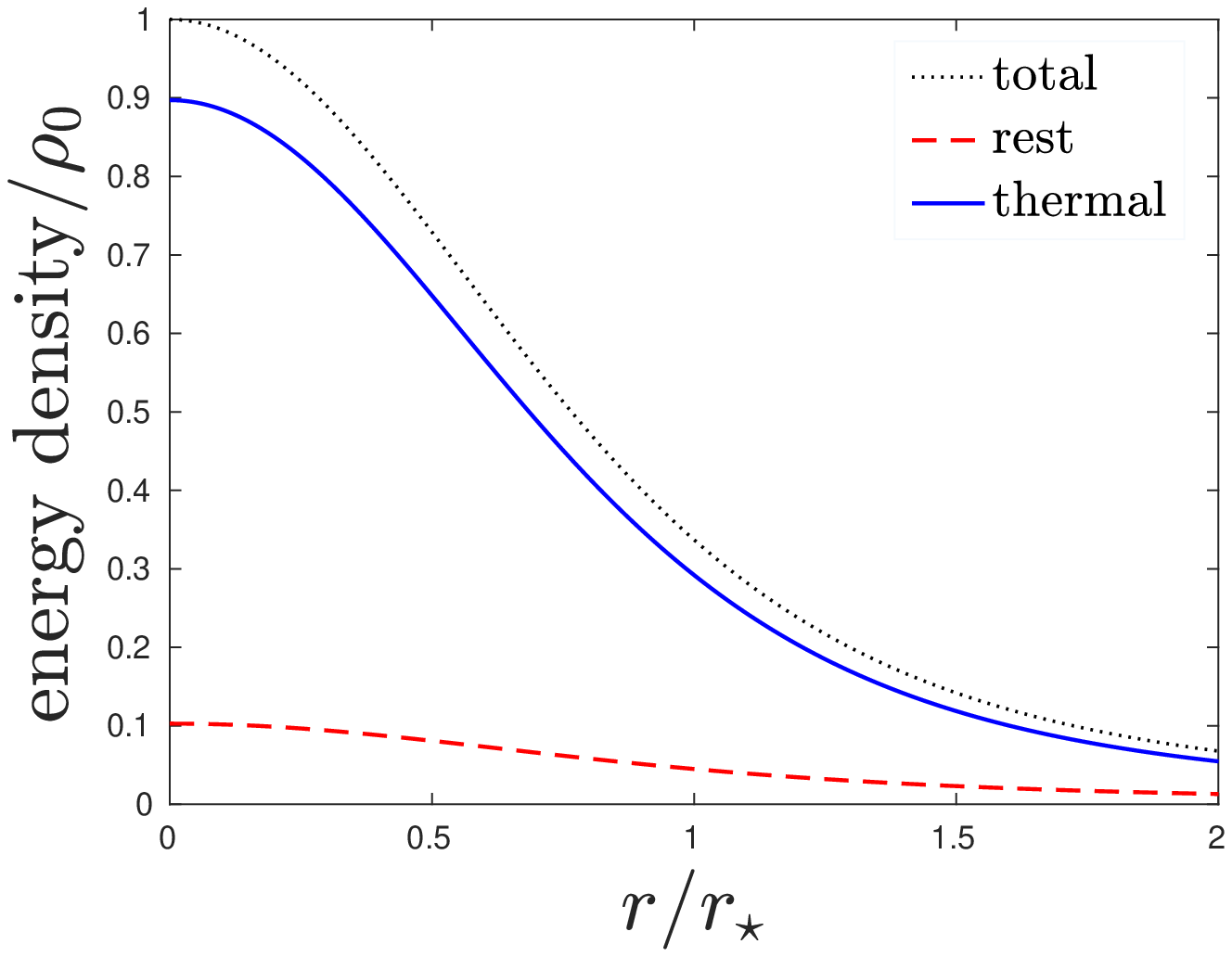}  }
		\caption{The mass-energy density distribution of the equilibrium point $A$ (left panel) and ${\Delta}$ (right panel) for rest compactness $\xi=0.1$. At~$A$, the low-energy gravothermal instability sets in, whereas at ${\Delta}$, the high-energy gravothermal instability. In~the low energy case, the core is dominated by the rest mass energy, whereas in the high energy case, the core is dominated by thermal mass-energy. 
			\label{fig:rho}}
	\end{center} 
\end{figure}
\unskip

\begin{figure}[tb]
	\begin{center}
		\includegraphics[scale = 0.4]{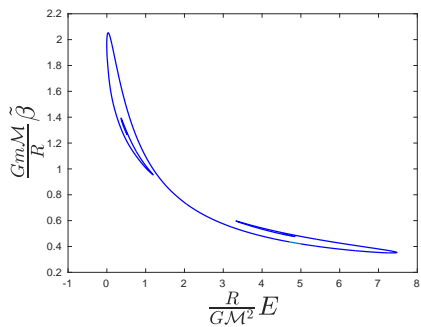}  
		\caption{The double spiral of the caloric curve $\tilde{\beta} = \tilde{\beta}(E)$ for rest compactness $\xi=0.25$ of the relativistic classical ideal gas reflecting the gravothermal instability. The~upper spiral is a manifestation of the low-energy gravothermal instability, and the lower spiral of the high-energy or relativistic gravothermal instability~\cite{Roupas_CQG_RGI_2015}. The~two spirals are connected with a stable series of equilibria. As~$\xi$ increases, the spirals approach each other and merge to a single point for $\xi = 0.35$. Beyond~this point, no equilibrium is~attainable.
			\label{fig:Spiral_double}}
	\end{center} 
\end{figure}
\unskip

In Figure~\ref{fig:spirals}a , the spiral of the caloric curve $\tilde{\beta} = \tilde{\beta} (E)$ is depicted, corresponding to the low-energy gravothermal instability. The~exact Antonov spiral is recovered for zero rest compactness $\xi \rightarrow 0$. As~$\xi$ increases the low-energy spiral is reformed, moving towards higher temperatures and energies. This means that as the sphere becomes more compact, and therefore relativistic effects more intense, the~stability domain is decreasing, and thus the system gets destabilized. This relativistic generalization of gravothermal catastrophe occurs to progressively higher minimum energies, which for~$\xi=0.1$ corresponds to point called $A$, as in Figure~\ref{fig:spirals}a.

As we move along this caloric curve from point $A$ to $B$, and then to higher energies and temperatures, the total mass-energy density profile becomes more homogeneous, because the dominating rest mass density gets more homogeneously distributed. This is manifested by a decrease of the density contrast $\ln\rho_0/\rho(R)$. However, as~temperature rises, the thermal energy density profile continuously steepens due to the Tolman--Ehrenfest effect---the weight of heat, at some point, denoted ${\Sigma}$ in Figure~\ref{fig:spirals}c---the~gravitation of thermal energy takes over its outward pointing, stabilizing, pressure effect. From~that point on, the~density profile becomes steeper, i.e.,~the density contrast increases, and~thermal mass gravity dominates over rest mass~gravity.

At sufficiently high energies, there appears a second spiral, the~high-energy one. At~the point of maximum energy, denoted ${\Delta}$ for $\xi=0.1$ in Figure~\ref{fig:spirals}b, the high-energy or relativistic gravothermal instability sets in. Figure~\ref{fig:spirals}c,d shows that this instability is similar to gravothermal catastrophe, in that it occurs as the system passes from negative to positive specific heat and not the other way around. This indicates that similarly to gravothermal catastrophe, in~the high-energy gravothermal instability a self-gravitating core with negative specific heat forms and decouples from the rest of the system. A~heat transfer from the core to the halo, i.e.,~a core--halo structure, leads to a runaway effect as the halo acquires positive specific heat; likewise, the whole system does in the unstable domain. However, the~big difference is shown in Figure~\ref{fig:rho}a,b. The~core at the onset of the high-energy gravothermal instability is dominated by thermal energy density and not rest mass-energy density, which is completely opposite to Antonov instability. The~system collapses under the weight of its own heat. The~temperature gradient from the core to the outer regions, formed at the onset of the instability, causes the self-gravitating (negative specific heat) core to heat up further, and thus become even heavier because it accumulates more heat. Thus, the~system becomes destabilized, as its energy is increasing and not decreasing, unlike the Antonov case. This is evidenced in Figure~\ref{fig:spirals}b. 

Both spirals, together with the stable branch connecting them, are shown in a single diagram in Figure~\ref{fig:Spiral_double} for $\xi=0.25$. As~the rest compactification is increasing, i.e.,~relativistic effects become more intense, the~relativistic spiral moves along lower energies, i.e.,~the stable domain gets smaller. This destabilization adds up to the destabilization caused by the low-energy spiral. The~two spirals approach each other with increasing $\xi$, and finally merge to a single point for
\begin{equation}
	\xi_\text{max} = 0.3529.
\end{equation}

This is an ultra-maximum limit of rest compactness. No static, stable, relativistic classical ideal gas can exist with rest mass to radius ratio higher than $\xi_\text{max}$. Thus, for~stable equilibria, it always holds that
\begin{equation}\label{eq:xi_max}
	\frac{2GmN}{Rc^2} < 0.3529.
\end{equation}

\begin{figure}[tb]
	\begin{center}
		\subfigure[Critical radius]{\label{fig:Rcr_micro}
			\includegraphics[scale = 0.4]{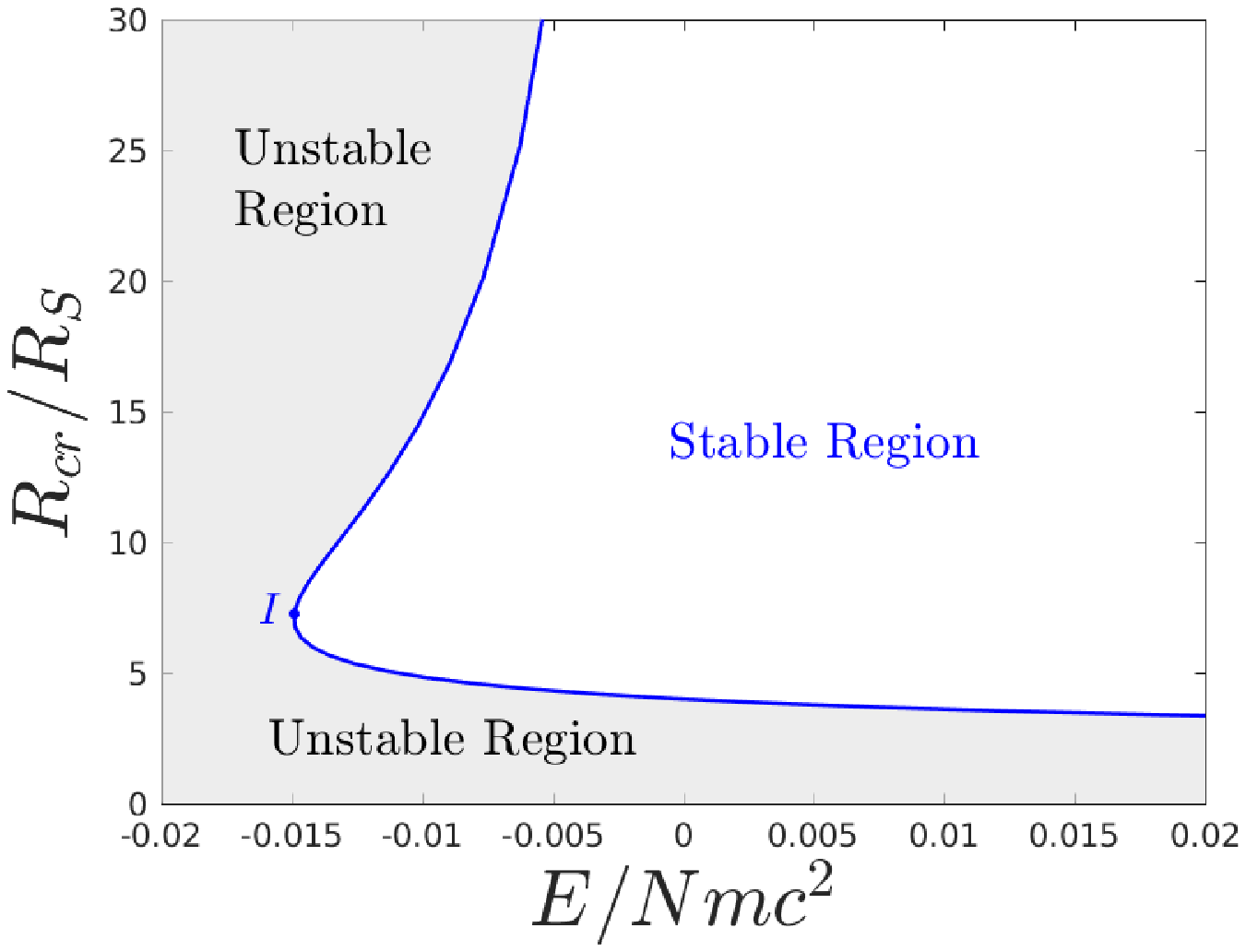}  }
		\subfigure[Critical mass and compactness]{\label{fig:Mcr_micro}
			\includegraphics[scale = 0.4]{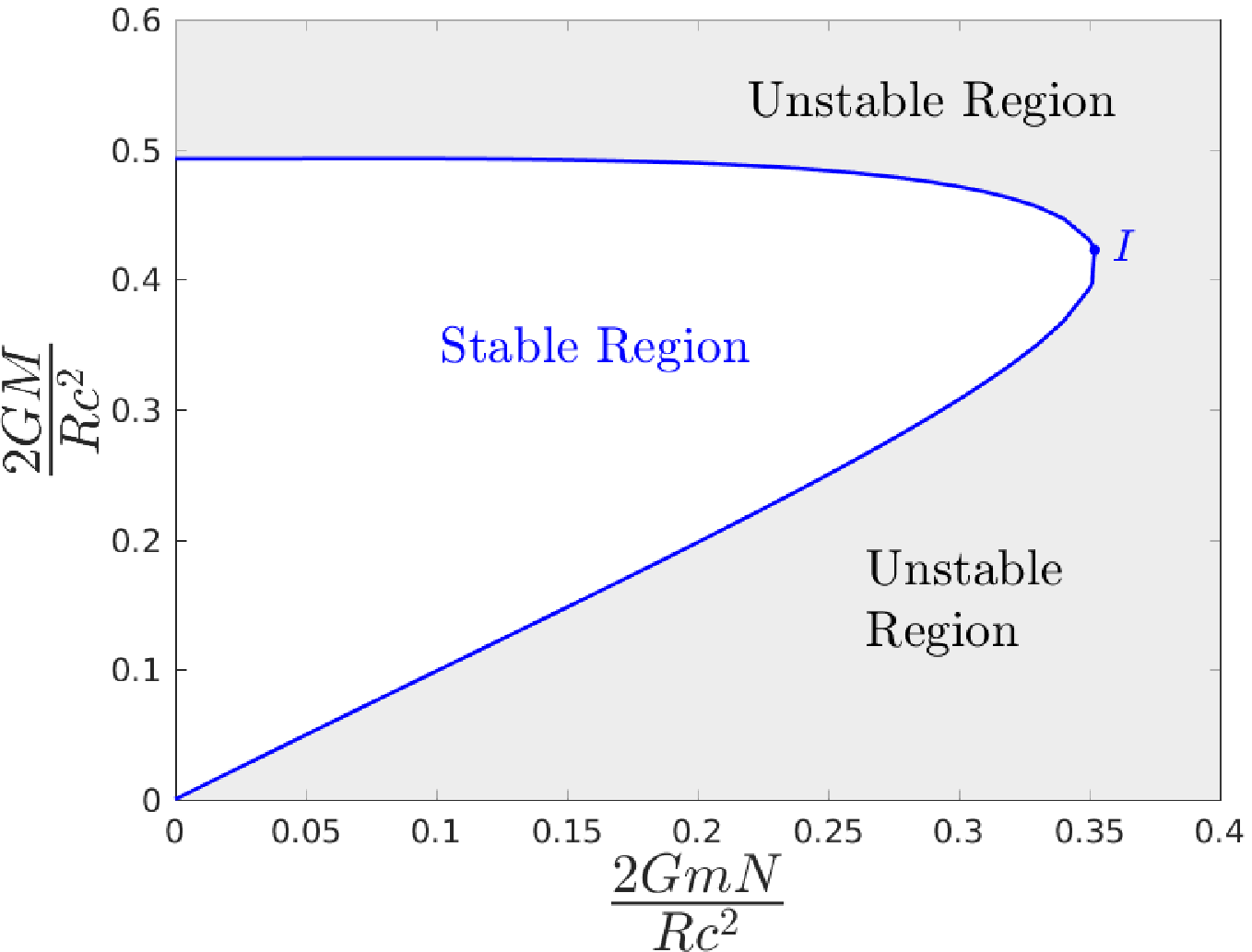}  }
		\caption{Left panel: The critical radius at which a gravothermal instability---for low or high energy---sets in with respect to the gravothermal energy. There appears a minimum gravothermal energy $E_\text{min} = -0.015 Nmc^2$ at point $I$ below which no equilibria exist. For~radii bigger than $I$, the~low-energy gravothermal instability sets in, while for radii smaller than $I$, the~relativistic gravothermal instability sets in. The~ultimate minimum radius is $2R_S$. Right panel: The critical compactness w.r.t. the rest compactness. Above~the maximum value, the high-energy gravothermal instability sets in, whereas below the minimum value, the low-energy one sets in. Point $I$ denotes the maximum possible rest compactness under~any conditions that equals $0.35$. The~maximum possible compactness is $0.5$.
			\label{fig:critical}}
	\end{center} 
\end{figure}

Let us now keep the energy fixed along with the number of particles and vary the radius of the system. This means we assume conditions of the microcanonical ensemble. If~the gravothermal energy
\begin{equation}
	E = Mc^2-Nmc^2
\end{equation}
is negative, but~sufficiently high so that stable equilibria do exist, there appear two critical radii, which delimit the stable domain shown in Figure~\ref{fig:critical}a. The~maximum radius is a manifestation of the low-energy instability and the minimum radius signifies the high-energy gravothermal instability. Therefore, in~the low-energy instability, the gas sphere becomes unstable when it becomes sufficiently large and not small. The~expansion causes the cooling of the system, which forces the rest mass towards the center in order to generate a pressure gradient strong enough to halt gravity. Above~the critical radius, a runaway heat transfer appears, with~direction from the core to the halo and the core collapses. On~the other hand if the radius is sufficiently decreased the resulted heating of the system forces thermal energy, and~not rest mass, to~concentrate on the center in order to generate a pressure gradient of a different origin in this case. Below~a critical radius, the high-energy gravothermal instability sets in and the system collapses with a similar core--halo mechanism. However, now the system becomes unstable when it becomes sufficiently small and not large. As~relativistic effects becomes more intense, i.e.,~when the absolute gravothermal energy, $|E|$ approaches a value closer to $Nmc^2$ and the~system gets destabilized, as the stable domain decreases. The~low-energy maximum radius decreases and the high-energy minimum radius increases. They merge at the ultra minimum gravothermal energy $E_\text{min}=-0.015Nmc^2$, represented by point $I$ in Figure~\ref{fig:critical}a. Therefore, for~stable equilibria, it always~holds that
\begin{equation}
	E > - 0.015 Nmc^2.
\end{equation}

This limit corresponds to the limit (\ref{eq:xi_max}). For~positive gravothermal energy $E$, there appears only the high-energy gravothermal instability, and thus the stable domain is bounded only by the minimum~radius.

In Figure~\ref{fig:critical}b, the critical compactness $2GM/Rc^2$ w.r.t. the rest compactness $2GmN/Rc^2$ is plotted. This plot may also be realized as the critical mass-energy w.r.t. the number of particles for a fixed radius $M_\text{cr} = M_\text{cr} (N)$. For~any rest compactness, there appear two critical energies: The~lower energy corresponds to the low-energy gravothermal instability and the higher energy to the high-energy gravothermal instability. The~maximum rest compactness given in (\ref{eq:xi_max}) corresponds to point $I$. The~total mass compactness  cannot be bigger than $0.5$.

Note that this analysis applies to the conditions of a microcanonical ensemble. This means that we assume adiabatic boundary conditions where no energy exchange between the system and its environment is allowed. In~a canonical ensemble, where the system is allowed to exchange heat with a reservoir, the~stability properties are completely different. This highlights the nonequivalence of ensembles in gravity (see, e.g., in~\cite{paddy}). In~the canonical case, both instabilities, in either the low or high energy regime, now called low-$T$ or high-$T$ isothermal collapse, respectively, occur below a minimum radius for fixed temperature~\cite{Roupas_CQG_RGI_2015}. However, in~the low-$T$ regime, the instability sets in below a critical temperature, and, in the high-$T$ one, above a critical temperature for fixed radius. In~the low-$T$ case, the decrease of temperature reduces the ability of the system to generate a pressure gradient, and, in the high-$T$ case, the increase of temperature enhances the concentration of thermal energy towards the center increasing its~gravitation.

Apparently, the~source of gravitational instability is in all of the above cases some heat transfer, either between subsystems of the system or the system and its environment: Gravitational instability manifests the universality of gravity and heat. The origin of gravitational instability, including quantum effects and interactions, will be discussed further in a separate~work.

\section{Core-Collapse~Supernova}

In a core-collapse supernova, the~collapse of the core of a massive star is initiated by the shift of nuclear statistical equilibrium (some modern reviews include~\cite{Janka_2012PTEP.2012aA309J,Janka_2012ARNPS..62..407J,Burrows_2013RvMP...85..245B}) when the nuclear fuel is exhausted. During~the collapse, the~core is heated up. Part of this thermal energy is consumed to dissociate heavy nuclei to nucleons with parallel emission of energetic neutrinos, while electron capture by protons enriches the core with neutrons. Neutrinos become trapped inside the core at densities $\sim$~10$^{12}$ gr/cm$^3$, and the collapse of the core may be halted at densities above the normal nuclear density $\sim 2.7 \cdot$ 10~$^{14}$gr/cm$^3$ with a bounce. At~the bounce, the core consists of an ultra-hot $\gtrsim 50 MeV$ nucleon gas, dominated by neutrons, and~trapped neutrinos of energy 100--300~MeV.

If the core at the time of the bounce lies in the unstable domain of Figure~\ref{fig:critical}b, which defines the stable domain in compactness space, it will be subject to the high-energy gravothermal instability and will not be able to halt collapse.
The temperature corresponding to the ultra maximum limit, point $I$ of Figure~\ref{fig:critical}b, of~relativistic gravothermal instability is $k\tilde{T}$ = 0.19 mc$^2$, that is, 178 MeV for~neutrons.

Note that the physics of the relativistic gravothermal instability, does not, qualitatively, depend on the equation of state. It is natural to expect that will persist for any equation of state. This is due to the universality of the effects of gravity and heat. Regarding our case of interest, namely, the quantum Fermi gas, this is proved in~\cite{Roupas_2015PRD,2019CQGra..36f5001R}. 

Note, in addition, that a system undergoing a gravothermal instability will be subject to heat transfer from a newly formed core to the outer regions, the~``halo''. Due to its negative specific heat, the core will get hotter and contracted. Such a phenomenon resembles the implosion--explosion structure of a supernova. If,~at some point, the temperature and compactness values allow for quantum degeneracy pressure to halt the collapse, the system will form a protoneutron star. In~this case, the core-collapse supernova may be viewed as a microcanonical gravitational phase transition~\cite{Alberti+Chavanis_2018,2019CQGra..36f5001R} from the initial gaseous (gravitational) phase of the massive star to the collapsed (gravitational) phase of the protoneutron star. This may only occur if the system is subject to the low-energy gravothermal instability, i.e.,~if at the onset of instability it lied below the lower line of Figure~\ref{fig:critical}b. 

\section{Conclusions}

This paper focused on two fundamental properties of matter: the~ability to move and to gravitate. The aim of this paper was to investigate how these phenomena intervene with respect to the stability of systems containing material particles that present only these phenomena. Such a system is called the relativistic, classical, self-gravitating ideal~gas. 

Although random movement, namely, thermal energy, naturally favors a disordered, homogeneous state, the~intriguing, universal character of gravity intervenes, and I propose that there exists a threshold beyond which the heating of the system does not homogenize, but steepens, the total mass-energy density. This is because the thermal energy gravitates. At~a critical point of maximum energy and minimum radius, an instability sets in. A~self-gravitating core with negative specific heat, dominated by the gravitation of thermal mass, decouples from the outer regions and collapses similarly to Antonov instability. A~relativistic generalization of Antonov instability---the low-energy gravothermal instability---also occurs, but,~in this case, beyond critical points of minimum energy and maximum radius. As~the relativistic effects get more intense, i.e.,~the compactness of rest mass is increasing, the~caloric curve, which has the form of a double spiral, decreases in size. At~some point, it reduces to a point where the two types of gravothermal instability, at~low and high energy, merge, revealing that they are aspects of a single~phenomenon.

Figure~\ref{fig:critical}b depicts the stable domain outside which a gravothermal instability occurs. I argue that if the collapsing ultra-hot core formed during a core-collapse supernova lies inside the unstable domain of Figure~\ref{fig:critical}b at the time of the bounce (when it achieves densities $\sim2.7\cdot$~10$^{14}$ gr/cm$^3$), it will be subject to the relativistic gravothermal instability. It will not be able to stabilize itself and continue to collapse towards the formation of a black hole.
Finally, the~current results support the idea that the implosion--explosion structure of supernovae, i.e.,~the implosion of the core with parallel explosion of outer parts, reflects the core--halo structure of the low-energy gravothermal~instability.

\bibliography{ROUPAS_Weight_of_Heat_2019}
\bibliographystyle{unsrt}

\end{document}